\documentclass[10pt,journal,final,twocolumn,twoside]{IEEEtran}

\usepackage[dvips]{graphicx}
\usepackage{epsfig}
\usepackage[cmex10]{amsmath}
\usepackage{amssymb}
\usepackage{amsthm}
\usepackage{amsfonts}
\usepackage{bm}
\usepackage{cite}
\usepackage[svgnames]{xcolor} 
\usepackage{pstricks,pst-node,pst-plot,pstricks-add}
\usepackage[tight,footnotesize]{subfigure}
\usepackage{algpseudocode}
\usepackage{algorithm}
\usepackage{tikz}
\usepackage{url}

\graphicspath{{figs/}}

\input{mohammad.def}

\begin{document}

	\title{Private Shotgun DNA Sequencing: A Structured Approach}
	
	\author{Ali Gholami, Mohammad Ali Maddah-Ali, and Seyed Abolfazl Motahari%
		\thanks{Ali Gholami is with the Department of Electrical Engineering, Sharif University of Technology, Tehran, Iran. (email:ali.gholami@ee.sharif.edu). Mohammad Ali Maddah-Ali is with Nokia Bell Labs, Holmdel, New Jersey, USA. (email:mohammad.maddahali@nokia-bell-labs.com). Seyed Abolfazl Motahari is with the Department of Computer Engineering, Sharif University of Technology, Tehran, Iran. (email:motahari@sharif.edu)}%
	}

	\maketitle

	\begin{abstract}

	DNA sequencing has faced a huge demand since it was first introduced as a service to the public. This service is often offloaded to the sequencing companies who will have access to full knowledge of individuals' sequences, a major violation of  privacy. 
	To address this challenge, we propose a solution, which  is based on separating the process of reading the fragments of sequences, which is done at a sequencing machine, and assembling the reads, which is done at a trusted local data collector. To confuse the sequencer, in a pooled sequencing scenario, in which multiple sequences are going to be sequenced simultaneously, for each target individual, we add fragments of one non-target individual, with a known DNA sequence at the data collector.  
	Then coverage depth of  the individuals, defined as the number of DNA fragments per DNA site, are selected proportional to the  powers of two. This layered structured solution allows us to ensure privacy, using only one sequencing machine, in contrast to our previous solution, where we relied on the existence of multiple non-colluding sequencing machines.
	\end{abstract}


	\section{Introduction}
	\label{sec:introduction}
	
Functionalities within the human body is coded in the DNA. The way cells evolve and form different tissues and limbs are highly correlated to the information stored in the genome. Human genome is a sequence of nucleotides chosen from the four member set $\{A,C,G,T\}$. The sequence in human genomes are very similar--more than 98 percent alike. What is mostly responsible for variations among human genomes are Single Nucleotide Polymorphisms (SNPs). In fact, an individual's genome can be uniquely characterized by its SNPs--that is called genotyping.

Having access to the genome sequence can benefit individuals for health care purposes both in diagnostic and therapeutic decision-making procedures \cite{phillips2014genomic}, \cite{van2013whole}, \cite{MedicalPress}. As a result, the usage of genetic testing services have risen massively in the past decade \cite{grosse2006clinical}, \cite{american2003american}, as well as genetic testing providers. As genomic data is becoming a leading part of health care procedures, concerns involving the privacy and confidentiality of this data have grown similarly \cite{anderlik2001privacy}, \cite{GenomeTheft}, \cite{heeney2011assessing}. The disclosure of this data can be maliciously used for example by insurance companies to increase the rates for particular diseases and drugs. Moreover, The disclosure of this information puts the information on the relatives in danger as well, due to the inherited similarities between family members \cite{knoppers2002genetic}, \cite{parker2004genetic}. Thus, accessing genomic data in one hand is useful in curing diseases and on the other hand its disclosure is a violation to the privacy of individuals \cite{kaye2012tension, lunshof2008genetic, malin2004not, lowrance2007identifiability}. There are a lot of papers addressing the issue of privacy in data exploration for genomic data. Some have used the concept of k-anonymity for providing data privacy, some have used differential privacy and others provided solutions by cryptographic methods \cite{sweeney2002k, aggarwal2005k, johnson2013privacy, fienberg2011privacy, kaye2009data, erlich2014routes, hasan2018secure, miller2018system, kantarcioglu2008cryptographic}. The objective in all those papers was to make sure no one'data is revealed in a published data set due to the process of sharing data for research purposes. In this paper, we have looked into the issue of privacy in a different way. The privacy is violated at the beginning of sequencing process, due to the access of the sequencing company to the sequence. Therefore, before we even disclose our data, the company knows our sequence.

The most popular method in sequencing the whole genome is shotgun DNA sequencing \cite{messing1981system}, \cite{venter1998shotgun}, \cite{motahari2013information}. In this method, the genome is broken into multiple fragments with various lengths. After that, a sequencing machine reads these fragments are assembles the \textit{reads} to build the whole sequence. Assembling algorithms available let the sequencing procedure to be both cost and time effective. It takes just a couple of days with a cost of less than 1000 dollars to sequence the genome, thanks to the existing sequencing machines. Also, to further reduce the costs and time, pooled sequencing can be used \cite{cutler2010pool}, \cite{cao2016combinatorial}, \cite{najafi2016fundamental}. In this methodology, rather than sequencing one individual, the genomes of a set of individuals are pooled together and sent to the sequencer. This will reduce the cost in comparison to the case in which these individuals sequenced the genome separately. Also, as wii be seen later on, the usage of pooled sequencing will benefit us in providing the privacy constraint.

Taking a deep look at the sequencing procedure, we realized that the sequencing process is itself a source of leakage for the sequence information. In this paper we introduce a scheme in which sequencing is possible while this kind of leakage is prevented and we will guarantee this privacy mathematically. In fact, we are going to sequence the genome of a set of individuals, using a sequencing machine, while limiting the knowledge received by the sequencer as desired. We first mention that the sequencing process consists of two phases. First is the \textit{reading phase} in which the sequencer reads the received fragments; i.e. determines the sequence of nucleotides in each fragment. Second is the \textit{processing phase} where a machine called \textit{data collector}, using the received reads, assembles the sequence of each individual. We aim at separating the two phases to provide privacy. In fact, we will introduce a methodology in which the sequencer is unable to do the processing phase while the data collector has the ability. In other words, the reading phase which needs high tech machines is outsourced, and the processing phase which is computational is done on a trusted local machine. To separate the two phases, we should make sure the data collector has more information in comparison to the sequencer. One of the ideas used in that direction is the usage of a set of individuals which their genome sequence is known a--priori to the data collector and unknown to the sequencer. the other idea is to use the finite field addition. Briefly, if we have two binary random variables and one of them has a uniform distribution, their summation in binary field reveals no new information of the non-uniform random variable; i.e. having the output of this summation, does not change the distribution of the random variable in comparison to the prior distribution. With these two ideas, we are going to limit the information leakage at the sequencer as desired, while letting the data collector to reconstruct the sequences.

This problem is conceptually connected to the Shamir sharing scheme \cite{shamir1979share}. In this scheme, a secret is partitioned to multiple parts, and each part is stored in a data base. This partition is done in such a way that with a subset of the data bases, the secret is reconstructed. In fact there is a threshold for the number of data bases where any subset with the number of data bases equal or more than that, can reconstruct the secret, and any subset with the number of data bases less than that threshold, receives no information about the secret \cite{gordon2006rational}. Based on this solution, there are many works providing solutions \cite{halpern2004rational}, \cite{arbogast2018parallelizing}. 

The rest of paper is organized as follows. The problem setting is provided in Section \ref{sec2}. In Section \ref{sec3}, an achievable scheme is introduced with the corresponding results. In Section \ref{sec4}, a generalized version of the scheme is introduced with the resulting theorems and Section \ref{sec6} concludes the paper and introduces some future steps.

	\section{Problem Setting} \label{sec2}
	
	We propose an architecture in which there is a trusted data collector and a sequencing machine (i.e. sequencer). also, there is a set of individuals that want their genome to be sequenced privately, without leaking the sequence data to the sequencer. There are $M\in \mathbb{N}$ individuals in this set and they are labeled from $0$ to $M-1$. The data collector has the duty to gather the genomes of the individuals in the set and pool their fragments (the genome is sheared to fragments with various sizes) together and send this pool to the sequencer. Then, the sequencer will read these fragments (\textit{reading phase}) and reports the resulting \textit{reads} to the data collector. At last, the data collector, using the set of reads, assembles the sequences for all individuals (\textit{processing phase}) and reports the results to them. To provide privacy, unlike conventional methods, we aimed at separating the reading phase with the processing phase. In fact, the sequencer has the duty to do the reading phase and the data collector is used for the processing phase. Our objective for privacy is to guarantee that the processing phase can not be done in the sequencer.
	
	To separate the two phases, we should create an information gap between the sequencer and the data collector. To do this, we use another set of individuals which their sequences are known before hand to the data collector but unknown to the sequencer. The genomes of this set of individuals are also collected by the data collector and their fragments are added to the pool. This set is of size $K\in\mathbb{N}$ and the individuals are labeled from $0$ to $K-1$ and are called \textit{known individuals}. The previous set, which the aim is to sequence their genomes, are called \textit{unknown individuals}. 
	
	We referred to SNPs earlier as the main source of difference between human genomes. Although there are four types of nucleotides, two of them can occur in every SNP position for all individuals, and this binary set in every position is known a--priori for the population. Also, for each SNP position, the allele occurring with more frequency in the population is called the major allele and the one occurring with less frequency is called the minor allele. Considering this, the sequence of every individual can be characterized by a vector in $\{0,1\}^N$ where $N\in\mathbb{N}$ is the total number of SNPs and $0$ and $1$ represent the minor and major alleles, respectively. Moreover, we define the matrix $\mathbf{X}$ which contains the random variable $X_{m,n}\in\{0,1\}$ in its row $m$ and column $n$ that indicates the allele for unknown individual $m$ in SNP position $n$. Similarly, the matrix $\mathbf{Y}$ is defined for the known individuals. Keep in mind that the entries in $\mathbf{X}$ are unknown both at the sequencer and the data collector, but the entries in $\mathbf{Y}$ are unknown to the sequencer and known to the data collector, leading to an information gap between these two.
	
	Let $\mathcal{F}_{m,n}$ and $\tilde{\mathcal{F}}_{k,n}$ denote the set of fragments containing SNP position $n\in\mathbb{N}$ for the unknown individual $m$ and known individual $k$ respectively. The data collector sends the set of fragments $\bigcup\limits_{m=1}^{M}\bigcup\limits_{n=1}^{N} \mathcal{F}_{m,n}+\bigcup\limits_{k=1}^{K} \bigcup\limits_{n=1}^{N} \tilde{\mathcal{F}}_{k,n}$ to the sequencer (see Fig.~\ref{fig:boat1}). Let us define the random variables $\alpha_{m,n} \triangleq |\mathcal{F}_{m,n}|$ and $\tilde{\alpha}_{k,n} \triangleq |\tilde{\mathcal{F}}_{k,n}|$ as the coverage depth for SNP position $n$ for the unknown individual $m$ and known individual $k$, respectively. Note that in the sequencing process, from each individual, there are a number of genomes provided for the data collector, so for most regions in the genome for one individual, there are multiple fragments containing the region. The sequencer reads each SNP with a probability of error. As will be seen later, to lower the effect of reading error caused by the sequencer, we should increase the coverage depth. The set of reads sent to the data collector by the sequencer is denoted by $\mathcal{R}$.
	
	\begin{figure}
		\begin{center}
			\begin{tikzpicture}
			\draw (2.75,1) rectangle (4.25,0)node [pos=.5] {Seq.};
			\draw (3.5,-1.5) circle (6.5 mm) node {D. C.};
			\draw (-0.75,-4) rectangle (0.75,-3)node [pos=.5] {\small U. Ind. $1$};
			\node at (1,-3.5) {$\cdots$};
			\draw (1.75,-4) rectangle (3.25,-3)node [pos=.5] {\small U. Ind. $M$};
			\draw (3.75,-4) rectangle (5.25,-3)node [pos=.5] {\small K. Ind. $1$};
			\node at (5.5,-3.5) {$\cdots$};
			\draw (6.25,-4) rectangle (7.75,-3)node [pos=.5] {\small K. Ind. $K$};
			\draw [thick, ->] (3.5,-8.5 mm) -- (3.5,0);
			\draw [thick, ->] (0,-3) -- (29.37 mm,-18.25 mm);
			\draw [thick, ->] (2.5,-3) -- (31.75 mm,-20.63 mm);
			\draw [thick, ->] (4.5,-3) -- (38.25 mm,-20.63 mm);
			\draw [thick, ->] (7,-3) -- (40.63 mm,-18.25 mm);
			\end{tikzpicture}
		\end{center}
		\caption{The block diagram of the proposed scheme in stage 1. First, the fragments of some individuals (known and unknown) are collected by the data collector, then they are pooled and sent to the sequencers.}
		\label{fig:boat1}
	\end{figure}
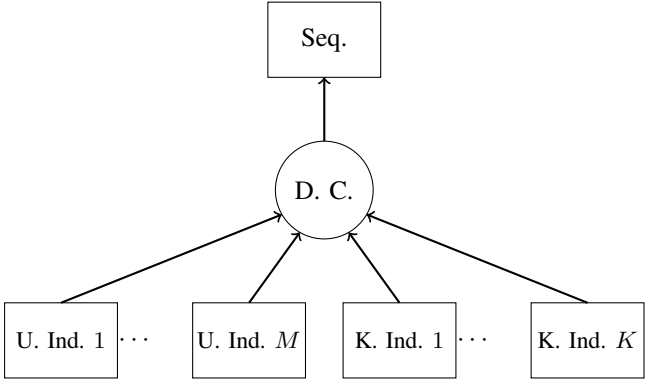
	
	\begin{figure}
		\begin{center}
			\begin{tikzpicture}
			\draw (2.75,1) rectangle (4.25,0)node [pos=.5] {Seq.};
			\draw (3.5,-1.5) circle (6.5 mm) node {D. C.};
			\draw (0,-4) rectangle (1.5,-3)node [pos=.5] {\small U. Ind. $1$};
			\draw (2.5,-4) rectangle (4,-3)node [pos=.5] {\small U. Ind. $2$};
			\node at (4.75,-3.5) {$\cdots$};
			\draw (5.5,-4) rectangle (7,-3)node [pos=.5] {\small U. Ind. $M$};
			\draw [thick, ->] (3.5,0) -- (3.5,-8.5 mm);
			\draw [thick, ->] (3.5,-21.5 mm) -- (3.25,-3);
			\draw [thick, ->] (30.4 mm,-19.6 mm) -- (0.75,-3);
			\draw [thick, ->] (39.6 mm,-19.6 mm) -- (6.25,-3);
			\draw (5.5,-8.5 mm) rectangle (9,-21.5 mm)node [pos=.5]
			{
				\begin{tabular}{cc}
				Sequence information \\
				of known individuals \\
				\end{tabular}
			};
			\draw [thick, ->] (5.5,-1.5) -- (41.3 mm,-1.5);
			\end{tikzpicture}
		\end{center}
		\caption{In stage 2, each sequencer sends the results of the reads to the data collector and then using the information of the known individuals, it will process the data and assemble the sequences of the unknown individuals.}
		\label{fig:boat2}
	\end{figure}
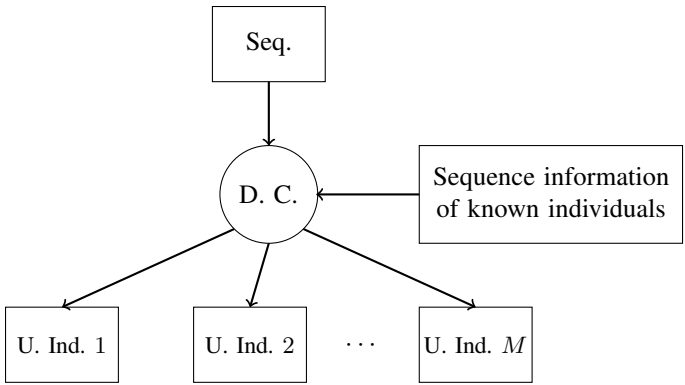
	
	Sequencers have errors in reading bases.  
	The probability of error in reading a SNP in a fragment is assumed to be constant across all sequences and for all SNPs and is denoted by  $\eta \in (0,1)$.
	More precisely, in the sequencer, for a fragment of an individual, and in a SNP, the probability that a $1$ is read $0$ or vice versa is $\eta$, independent of the individual, the fragment, and the SNP. 
	
	Having $\mathbf{Y}$ as a side-information, the data collector maps $\mathcal{R}$ to the matrix $\hat{\mathbf{X}} \in \{0, 1\}^{M \times N}$ 
	using a function $\phi$, i.e.   
	\begin{align*}
	\hat{\mathbf{X}} = \phi \left(\mathbf{Y},  \mathcal{R}\right),
	\end{align*}
	where $\hat{\mathbf{X}}$ refers to an estimate of the matrix of SNPs for unknown individuals $\left(\mathbf{X}\right)$.
	
	The proposed scheme should be such that the following two conditions are satisfied:
	\begin{itemize} \label{furmulbandi}
		\item \emph{Reconstruction Condition:}
		Let  $\mathbf{x}_n$ and $\hat{\mathbf{x}}_n$ denote the column $n$ of the matrix $\mathbf{X}$ and $\hat{\mathbf{X}}$ respectively. 
		The reconstruction condition requires that the inequality below hold for any given  $\epsilon \in (0,1)$:
		\begin{equation} \label{rec}
		\mathbb{P}(\hat{\mathbf{x}}_n\ne \mathbf{x}_n)\leq \epsilon,\quad\forall n\in [N].
		\end{equation} 
		$\epsilon$ is referred to as the \emph{accuracy level} and is a design parameter. 
		
		\item \emph{Privacy Condition:} 
		For privacy to be held, we want the distribution of $X_{m,n},   m \in [0:M-1], n \in [N]$ remains almost the same before and after reading the fragments. To be precise,  the privacy condition requires that the following inequality hold for any given  $\beta \in (0,1)$:
		\begin{equation} \label{pri}
		\frac{I\left(  X_{m,n},   m \in [0:M-1], n \in [N]   ;\mathcal{R}\right)}{MN}\leq\beta.
		\end{equation}
		$\beta$ is referred to as the \emph{privacy level} and is a design parameter. 
	\end{itemize}
	
	In the following section we will introduce a proposed scheme that satisfies the two conditions simultaneously.
	
	\section{Structured Achievable Scheme with Constant Coverage Depth} \label{sec3}
	In this section, we propose a scheme to satisfy both the reconstruction~\eqref{rec} and privacy~\eqref{pri} constraints. We have two assumptions in our scheme:
	\begin{itemize}
		\item \emph{Assumption 1:} Every fragment is short enough to contain no more than one SNP.
		\item \emph{Assumption 2:} Every fragment is long enough that can be correctly mapped to the reference genome, i.e. we can identify exactly from what region of the genome sequence they came from.
	\end{itemize}
	
	These two assumptions are realistic. We should keep in mind that there are approximately 3.3 million SNPs in the human genome. Comparing to the 3 billion length of the whole genome, it is concluded that the average distance between two SNPs is roughly 1000 base pairs \cite{shen2013comprehensive}. Moreover, using short read alignment algorithms like Bowtie \cite{langmead2009ultrafast}, it is possible to assemble reads of length in the order of a couple of hundreds. Thus using such algorithms, and choosing the fragments lengths to be about few hundreds, both assumptions are valid simultaneously.
	
	In the proposed achievable scheme, we focus on the case where $S=M$. In cases where $M$ is greater than $S$, we partition the set of individuals into groups of size $S$ and use this scheme for each group separately. In this paper, we propose a specific assignment scheme for the coverage depth parameters. In the proposed solution, named \emph{structured scheme}, for $\forall m\in [0:M-1]$,  $\forall k\in [0:K-1]$, and $\forall n\in [N]$ we have
	\begin{align}
	\alpha_{m,n}=2^m\alpha_0,
	\end{align}
	and
	\begin{align}
	\tilde{\alpha}_{k,n}=2^k\alpha_0.
	\end{align}
	where $\alpha_0\in \mathbb{N}$. Also, entries in $\mathbf{X}$ have prior probabilities following the major allele frequencies and entries in $\mathbf{Y}$ have uniform prior probabilities. 
	
	Keeping the coverage depth variables exactly as introduced in the above equations is practically impossible. They are actually random variables. Analyzing the random case is rather complicated. To have a better understanding of the problem and make the analysis tractable, in this section, we consider the constant case and later in Section Blah, we generalize the results to the case of random coverage depths.  

	First, we introduce the main results. Then we derive the mathematical models in the data collector and the sequencers in Subsections \ref{mathmode1} and \ref{mathmode2}, respectively.  We rely on these models to prove the main results in Subsections \ref{proof1} and \ref{proof2}. At last, we discuss the results in Subsection \ref{dis}.
	
	The following theorem provides a sufficient condition for the reconstruction condition to hold.
	\begin{theorem} \label{thm1}
		In the structured scheme with constant coverage depth and reading probability of error of $\eta$, the reconstruction condition~\eqref{rec} is satisfied if  
		\begin{align} \label{first}
		\frac{\alpha_0}{2^{M+1}-2}\geq\frac{8\eta(1-\eta)}{(1-2\eta)^2}\ln\left(\frac{1}{\epsilon}\right).
		\end{align}
	\end{theorem}
	
	The following theorem provides a sufficient condition for the privacy condition to hold.
	\begin{theorem} \label{thm2}
		In the structured scheme with constant coverage depth, the privacy condition~\eqref{pri} is satisfied if
		\begin{align} \label{second}
		M\geq\frac{1}{\beta}.
		\end{align}
	\end{theorem}
	
	The main message of these results is that we can choose the parameters of the proposed scheme such that both conditions are satisfied, simultaneously. In other words, these theorems confirm that the separation of the reading phase and the processing phase together with adding known individuals and by adjusting coverage depths, offers enough flexibility to satisfy both conditions at the same time; based on~\eqref{second}, $M$ is chosen, and using~\eqref{first}, $\alpha_0$ is set.
	
	\begin{example}
		If we assume the values $\eta=\epsilon=\beta=0.01$, then based on Theorem \ref{thm2}, we can have $M\geq100$ and based on Theorem \ref{thm1} for $M=100$, $\alpha_0\simeq 9.6\times 10^{29}$ (or greater). Also for $\eta=\epsilon=\beta=0.1$, we get $M=10$ and $\alpha_0\simeq 5300$. For another example if we assume the values $\eta=0.01$, $\epsilon=0.001$, and $\beta=0.1$, we will have $M=10$ and $\alpha_0\simeq 1166$.
	\end{example}
	
	\subsection{Mathematical Model in Data Collector in the Structured Scheme} \label{mathmode1}
	For any SNP position $n\in [N]$, the data collector should be able to estimate the vector $\textbf{x}_n=\left[X_{0,n},X_{1,n},\cdots,X_{M-1,n}\right]$.
	
	In this subsection, we seek for the model that the data collector observes in SNP position $n$. We will show that the data collector receives $G_n$ as
	\begin{align} \label{maindc}
	G_n=\sum_{m=0}^{M-1}2^mX_{m,n}+Z_n,
	\end{align}
	in which $Z_n\sim\mathcal{N}\left(0,\sigma^2\right)$ where
	\begin{align} \label{sigma2}
	\sigma^2\triangleq \frac{2^{M+1}-2}{\alpha_0}\frac{\eta (1-\eta)}{(1-2\eta)^2}.
	\end{align}
	
	To obtain this model, we should keep in mind that the fragments have no tags and the data collector and sequencer both do not know the corresponding individual which every fragment belongs to. Therefore, when the data collector receives the read fragments from sequencer, the only information it gets is the number of major (or minor) alleles in every position $n\in [1:N]$. Consequently, the data collector receives the following summation
	\begin{align} \label{12}
	\sum_{m=0}^{M-1} \sum_{i=1}^{2^m\alpha} \left(\tilde{X}_{m,n,i}+\tilde{Y}_{m,n,i}\right),
	\end{align} 
	in which $\tilde{X}_{m,n,i}$ and $\tilde{Y}_{m,n,i}$ are noisy versions of $X_{m,n}$ and $Y_{m,n}$ respectively, due to the reading error caused by the sequencer. Also, recall that the data collector knows the sequence of known individuals a priori, i.e. it knows the value for all $Y_{m,n}$. Let us assume these values are $Y_{m,n}=y_{k,n}$. Therefore we have
	\begin{align}
	\tilde{X}_{m,n,i} &=\left\{
	\begin{array}{ll}
	X_{m,n},   & \mathrm{w.p.}\quad 1-\eta \\‎ 
	1-X_{m,n},  & \mathrm{w.p.}\quad\eta,
	\end{array}\right.\label{XXX}\\
	\tilde{Y}_{m,n,i} &=\left\{
	\begin{array}{ll}
	y_{m,n},   & \mathrm{w.p.}\quad 1-\eta \\‎ 
	1-y_{m,n},  & \mathrm{w.p.}\quad\eta.
	\end{array}\right.\label{YYY}
	\end{align}
	Note that the $i$ index refers to the \textit{read} number. After scaling~\eqref{12} and subtracting $\sum_{k=0}^{M-1}y_{k,n}$ and $2^{M+1}\frac{\eta}{1-2\eta}$,~\eqref{12} can be written as
	\begin{align} \label{main}
	G_n&=\frac{1}{\alpha_0 (1-2\eta)}\left(\sum_{m=0}^{M-1} \sum_{i=1}^{2^m\alpha_0} \left(\tilde{X}_{m,n,i}+\tilde{Y}_{m,n,i}\right)\right)\nonumber\\
	&-\sum_{k=0}^{M-1}y_{k,n}-2^{M+1}\frac{\eta}{1-2\eta}.
	\end{align}
	Note that subtracting $\sum_{k=0}^{M-1}y_{k,n}$ in the above equation is fine, because of the full knowledge of matrix $\mathbf{Y}$ is available at the data collector. 
	
	To follow, we derive the parameters of the random variable $\tilde{X}_{m,n,i}$ on the condition of knowing $X_{m,n}$. Based on~\eqref{XXX} we have
	\begin{align}
	\mathbb{E}\left(\tilde{X}_{m,n,i}|X_{m,n}\right)&=X_{m,n} (1-\eta)+(1-X_{m,n})\eta\nonumber\\
	&=(1-2\eta) X_{m,n}+\eta\\
	\mathrm{Var}\left(\tilde{X}_{m,n,i}|X_{m,n}\right)&=\mathbb{E}\left(\left(\tilde{X}_{m,n,i}\right)^2|X_{m,n}\right)\nonumber\\
	&-\left(\mathbb{E}\left(\tilde{X}_{m,n,i}|X_{m,n}\right)\right)^2\nonumber\\
	&=\left(X^2_{m,n}(1-\eta)+\left(1-X_{m,n}\right)^2(\eta)\right)\nonumber\\
	&-\left((1-2\eta)X_{m,n}+\eta\right)^2\nonumber\\
	&=\eta (1-\eta),
	\end{align}
	in which the last inequality is valid for both possible values of $X_{m,n}$; i.e. $0$ and $1$. Using the MMSE estimate and orthogonality principle, we can write
	\begin{align} \label{model1}
	\tilde{X}_{m,n,i}=(1-2\eta)X_{m,n}+\eta+Z_{m,n,i},
	\end{align} 
	where $Z_{m,n,i}$ is a random variable with $\mathbb{E}(Z_{m,n,i})=0$ and $\textrm{Var}(Z_{m,n,i})=\eta (1-\eta)$. Also $Z_{m,n,i}$ and $X_{m,n}$ are uncorrelated.
	Consequently
	\begin{align} \label{model}
	\frac{1}{\alpha_0 (1-2\eta)}\sum_{i=1}^{2^m\alpha_0} \tilde{X}_{m,n,i}=2^mX_{m,n}+\frac{2^m\eta}{1-2\eta}+\frac{\sum_{i=1}^{2^m\alpha_0}Z_{m,n,i}}{\alpha_0 (1-2\eta)}.
	\end{align}
	Based on central limit theorem $\frac{\sum_{i=1}^{\alpha_0}Z_{m,n,i}}{\sqrt{\alpha_0}}$ converges in distribution to a normal distribution with zero mean and variance $\eta(1-\eta)$. Thus
	\begin{align}
	\frac{\sum_{i=1}^{\alpha_0}Z_{m,n,i}}{\alpha_0 (1-2\eta)}=\frac{1}{\sqrt{\alpha_0}(1-2\eta)}\frac{\sum_{i=1}^{\alpha_0}Z_{m,n,i}}{\sqrt{\alpha_0}}
	\end{align}
	converges in distribution to a normal distribution with zero mean and variance $\frac{\eta(1-\eta)}{\alpha_0(1-2\eta)^2}$. Thus, the last term in the right-hand side of~\eqref{model} converges to a normal distribution with zero mean and variance $2^m\eta(1-\eta)$. Similarly Using~\eqref{YYY}, we reach a similar equation.
	
	Consequently, using~\eqref{model}, we can rewrite~\eqref{main} as~\eqref{maindc}.

	\subsection{Mathematical Model in Sequencer in Structured Scheme} \label{mathmode2}
	Similar to the previous subsection, the sequencer receives the following summation in~\eqref{12}. The difference here with the previous subsection is that all individuals are unknown form the sequencer's view point. Therefore,
	\begin{align}
	\tilde{Y}_{k,n,i} &=\left\{
	\begin{array}{ll}
	Y_{k,n},   & \mathrm{w.p.}\quad 1-\eta \\‎ 
	1-Y_{k,n},  & \mathrm{w.p.}\quad \eta.
	\end{array}\right.\label{Y}
	\end{align}
	Yet, $\tilde{X}_{m,n,i}$ follows~\eqref{XXX}.
	
	Scaling the summation in~\eqref{12}, the sequencer receives $q_n$ defined as
	\begin{align}
	q_n\triangleq \frac{1}{\alpha_0 (1-2\eta)}\left(\sum_{m=0}^{M-1} \sum_{i=1}^{2^m\alpha_0} \left(\tilde{X}_{m,n,i}+\tilde{Y}_{m,n,i}\right)\right)
	\end{align}
	
	Taking similar steps as in the previous subsection, $q_n$ is written as
	\begin{align} \label{mains}
	q_n=\sum_{m=0}^{M-1}2^m(X_{m,n}+Y_{m,n})+\tilde{Z}_n,
	\end{align}
	where $\tilde{Z}_{n}\sim\mathcal{N}\left(0,\sigma^2\right)$ in which $\sigma^2$ is defined in~\eqref{sigma2}.

	\subsection{Proof of Theorem \ref{thm1}} \label{proof1}
	Having reached the mathematical model in the data collector in~\eqref{maindc}, we provide the proof of theorem \ref{thm1}.
	\begin{proof}
		Note that the value of the summation $\sum_{m=0}^{M-1}2^mX_{m,n}$ uniquely matches to a $\textbf{x}_n$ (in binary representation of it, each entry corresponds to a $X_{m,n}$ for different values of $m$). Therefore, our objective is to find the summation above. The probability of error in estimating the summation, based on~\eqref{maindc}, is simply upper bounded by
		\begin{align}
		\mathbb{P}(\text{error}) \leq Q\left(\frac{d_{\text{min}}}{2\sigma}\right).
		\end{align}
		Obviously, here $d_{\text{min}}=1$ due to the fact that $X_{m,n}$s are chosen from the set $\{0,1\}$. Thus
		\begin{align}
		\mathbb{P}(\text{error}) \leq Q\left(\frac{1}{2\sigma}\right)\leq\exp\left(\frac{-1}{8\sigma^2}\right),
		\end{align}
		in which $\sigma^2$ is defined in~\eqref{sigma2}.
		
	\end{proof}
	
	\subsection{Proof of Theorem \ref{thm2}} \label{proof2}
	Using the mathematical model in~\eqref{mains}, we are ready to provide the proof of theorem \ref{thm2}.
	\begin{proof}
		The fact is that for the sequencers, $\mathcal{R}$ is equivalent to $q_n$, $\forall n \in [N]$ because fragments contain just one SNP and are grouped based on their containing SNP position and in the group containing SNP position $n$, the information is stored in $q_n$. Thus we have
		\begin{align}
		\mathbb{P}\left(\mathbf{X}|\mathcal{R}\right)&=\prod_{n=1}^{N}\mathbb{P}\left(\mathbf{x}_{n}|q_n\right).
		\end{align}
		Recall that $\mathbf{x}_{n},$ $\forall n\in [N]$ denotes the column $n$ of $\mathbf{X}$. Due to independence of entries in $\mathbf{X}$, we have
		\begin{align}
		\mathbb{P}(\mathbf{X})=\prod_{n=1}^{N}\mathbb{P}(\mathbf{x}_{n}).
		\end{align}
		Based on the last two equalities
		\begin{align}
		I\left(\mathbf{X};\mathcal{R}\right)=\sum_{n=1}^{N}I\left(\mathbf{x}_n;q_n\right).
		\end{align}
		Thus, for privacy condition~\eqref{pri} to be satisfied, it is sufficient for every $n\in [N]$ to have
		\begin{align} \label{r3}
		\frac{I\left(\mathbf{x}_n;q_n\right)}{M}\leq \beta.
		\end{align}
		
		To begin, we define $Z_n$ as
		\begin{align} \label{Zdef}
		Z_n\triangleq \sum_{m=0}^{M-1}2^m(X_{m,n}+Y_{m,n}) 
		\end{align}
		It is concluded that the following Markov chain holds,
		\begin{align} \label{markov}
		\mathbf{x}_n\rightarrow Z_n\rightarrow q_n
		\end{align}
		Thus we have
		\begin{align} \label{dpi}
		I(\mathbf{x}_n;q_n)\leq I(\mathbf{x}_n;Z_n).
		\end{align}
		In what follows, we seek for $I(\mathbf{x}_n;Z_n)$. We have
		\begin{align} \label{entexpansion}
		I(X_{0,n},\cdots,X_{M-1,n};Z_n)&=H(Z_n)\nonumber\\
		&-H(Z_n|X_{0,n},\cdots,X_{M-1,n}).
		\end{align}
		We expand $Z_n$ in binary formation
		\begin{align} \label{binaryexp}
		Z_n=(B_{M,n}b_{M-1,n}\cdots b_{0,n})_2.
		\end{align}
		Consequently, the following equations hold
		\begin{align}
		X_{0,n}+Y_{0,n}&=2B_{1,n}+b_{0,n},\\
		X_{1,n}+Y_{1,n}+B_{1,n}&=2B_{2,n}+b_{1,n},\\
		&\vdots\nonumber\\
		X_{M-1,n}+Y_{M-1,n}+B_{M-1,n}&=2B_{M,n}+b_{M-1,n},
		\end{align}
		in which in equation $i$, $B_{i+1}$ is the carry over of the left-hand summation in binary field. Equivalently we have
		\begin{align}
		b_{0,n}&=X_{0,n}\oplus Y_{0,n},\label{1}\\
		b_{1,n}&=X_{1,n}\oplus Y_{1,n}\oplus B_{1,n},\label{2}\\
		&\vdots\nonumber\\
		b_{M-1,n}&=X_{M-1,n}\oplus Y_{M-1,n}\oplus B_{M-1,n}\label{3}.
		\end{align} 
		~\eqref{binaryexp} yields
		\begin{align}
		H(Z_n)=H(B_{M,n}b_{M-1,n}\cdots b_{0,n}).
		\end{align}
		We expand the right hand side of the above equality as
		\begin{align} \label{Hexpansion}
		H(B_{M,n}b_{M-1,n}\cdots b_{0,n})&=H(b_{0,n})+H(b_{1,n}|b_{0,n})\nonumber\\
		&+\cdots+H(B_{M,n}|b_{M-1,n}\cdots b_{0,n}).
		\end{align}
		Based on~\eqref{1} and the fact that entries of $\mathbf{Y}$ have uniform prior probabilities, $b_{0,n}$ has uniform distribution, so $H(b_{0,n})=1$. For $H(b_{1,n}|b_{0,n})$ we have
		\begin{align}
		H(b_{1,n}|b_{0,n})\geq H(b_{1,n}|b_{0,n}, B_{1,n})&\stackrel{(a)}{=}H(b_{1,n}|B_{1,n})\nonumber\\
		&\stackrel{(b)}{=}H(X_{1,n}\oplus Y_{1,n}),
		\end{align}
		which also results in $1$. Note that $(a)$ is resulted from the fact that $B_{1,n}$ is sufficient statistic for $b_{1,n}$. Also $(b)$ is resulted from~\eqref{2}. Similarly, all the terms in~\eqref{Hexpansion} result to $1$ except the last term. Therefore,
		\begin{align} \label{shalala}
		H(Z_n)&=H(B_{M,n}b_{M-1,n}\cdots b_{0,n})\nonumber\\
		&=M+H(B_{M,n}|b_{M-1,n}\cdots b_{0,n}).
		\end{align}
		Based on~\eqref{Zdef}, for the second term in the right hand side of~\eqref{entexpansion} we have
		\begin{align}
		H(Z_n|X_{0,n},\cdots,X_{M-1,n})&=H\left(\sum_{m=0}^{M-1}2^mY_{m,n}\right)\nonumber\\
		&=\sum_{m=0}^{M-1}H(Y_{m,n})=\sum_{m=0}^{M-1}1=M.
		\end{align}
		Using the last two equalities and~\eqref{entexpansion}, we have
		\begin{align} \label{leakage}
		I(\mathbf{x}_n;Z_n)=H(B_{M,n}|b_{M-1,n}\cdots b_{0,n})\leq 1.
		\end{align}
		The proof is complete.
		
	\end{proof}
	
	\subsection{Discussion} \label{dis}
	As it is seen from theorem \ref{thm1}, the minimum $\alpha_0$ needed to preserve the reconstruction condition, behaves exponential with $M$. $\alpha_0$ is a noise-resistance parameter and as it becomes larger, the ratio of the fragments containing false reads concentrate to the probability of error in the reading phase ($\eta$); that is why increasing $\alpha_0$ helps to eliminate the noise term in~\eqref{maindc}.
	
\begin{figure}
		\centering
		\hspace*{-0.9cm}
		\includegraphics[width=1\linewidth]{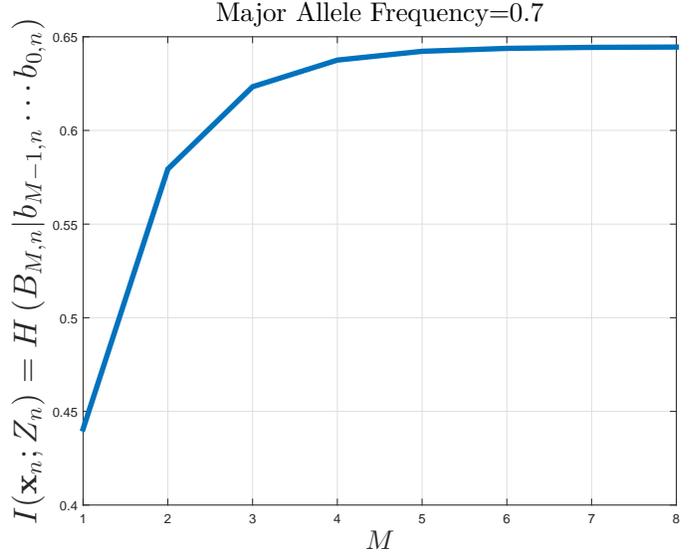}
		\caption{As seen in this figure, $I(\mathbf{x}_n;Z_n)$ is an increasing function with increasing $M$.}
		\label{ENT}
	\end{figure}

	Taking a deeper look at the procedure in the proof of Theorem \ref{thm2}, we realize that we have created the binary field addition in our scheme, as was desired. The bits $b_{i,n}$ that derive form~\eqref{1} to~\eqref{3}, are the result of binary field addition. The addition is for two random variables where one of them has uniform distribution, $Y_{i,n}$, and the other, $X_{i,n}$,follows the distribution of SNP position $n$. If the value of $b_{i,n}$ is given alone, the results reveals no new information about $X_{i,n}$. Thus these bits alone, are not leaking any information. So we have created this kind of addition, thanks to adjusting the coverage depth values. From~\eqref{maindc} it is concluded that the only bit leaking information in position $n$ is $B_{M,n}$ which means the binary field addition scheme is not working perfectly, but we should remember that the problem addressed in this paper has its limitations that we should adapt to. Interestingly, the maximum entropy of this bit is $1$ and this upper bound on the information leakage is independent of $M$. This aspect is very interesting and useful and results the average information leakage per bit to be at most $\frac{1}{M}$. Therefore by increasing $M$, this average decreases, so we can adjust $M$ so that we reach the desired $\beta$. Note that based on our simulations, $I(\mathbf{x}_n;Z_n)=H(B_{M,n}|b_{M-1,n}\cdots b_{0,n})$ is an increasing function of $M$ (see Figure \ref{ENT}) and tends to an ultimate value. So by increasing $M$, the information leakage per bit decreases with the rate of $
	\frac{1}{M}$, not more.
	
	\section{Structured Achievable Scheme with Random Coverage Depth} \label{sec4}
	
	In the previous section, we analyzed the problem for constant coverage depth; however, it is not a practical case because we do not have exact control on the number of fragments. In this section, we consider a more general case in which the coverage depth parameters are random variables. We assume them to be binomial variables and approximate them with normal distribution. Therefore, for $\forall n\in [N], \forall m\in [0:M-1]$, we have
	\begin{align}
	\alpha_{m,n}\sim \mathcal{N}\left(2^m\alpha_0, 2^m\sigma_{\alpha}^{2}\right).
	\end{align} 
	Similarly for $\forall n\in [N], \forall k\in [0:K-1]$, we have
	\begin{align}
	\tilde{\alpha}_{k,n}\sim \mathcal{N}\left(2^k\alpha_0, 2^k\sigma_{\alpha}^{2}\right).
	\end{align} 
	Due to the fact that coverage depths mostly have large values, we assumed that $\alpha_0\in\mathbb{N}$.
	
	As the previous section, we introduce the results hereunder. After that, the mathematical model and the estimation rule are introduced in Subsections \ref{mathmodel} and \ref{est}. Then, the proof of Theorem \ref{thm3} is provided in Subsection \ref{thm3proof}. Following them, we discuss the results in Subsection \ref{disc}.
	
	The following theorem provides a sufficient condition to satisfy the reconstruction condition.
	\begin{theorem}
		\label{thm3}
		In the all-but-one scheme, the reconstruction condition~\eqref{rec} is satisfied if:
		\begin{align} \label{eqeq}
		\alpha_0\geq\max\{e_1,e_2\},
		\end{align}
		where
		\begin{align}
		e_1\triangleq \frac{16\eta (1-\eta)}{(1-2\eta)^2}\left(2^{M+1}-2\right)\ln\left(\frac{1}{\epsilon}\right),
		\end{align}
		and
		\begin{align}
		e_2\triangleq \sqrt{16\sigma_\alpha^2\left(1+\frac{\eta^2}{(1-2\eta)^2}\right)\left(2^{M+1}-2\right)\ln\left(\frac{1}{\epsilon}\right)}.
		\end{align}
	\end{theorem}
	
	\begin{remark}
		For the privacy condition, Theorem \ref{thm2} is valid here as well. This will be discussed later in Subsection \ref{discuss2}.
	\end{remark} 
	
	\subsection{Mathematical Model in Data Collector in the Structured Scheme} \label{mathmodel}
	In this subsection, we will show that the information the data collector receives is the value in $G_n$ which is written as
	\begin{align} \label{sss}
	G_n=&\sum_{m=0}^{M-1} \left(2^m+\delta_{m,n}\right)X_{m,n}\nonumber\\
	+&\sum_{k=0}^{M-1} \left(2^k+\tilde{\delta}_{k,n}\right)y_{k,n} + Z_n
	\end{align}
	where $\delta_{m,n}$ and $\tilde{\delta}_{k,n}$ are normal random variables with zero mean and variance $2^m\sigma_1^2$ and $2^k\sigma_1^2$ respectively, where
	\begin{align} \label{sigma1}
	\sigma_1^2\triangleq\frac{\sigma_\alpha^2}{\alpha_0^2}.
	\end{align}
	Also, $Z_n\sim\mathcal{N}(0,\sigma^2)$ where
	\begin{align} \label{sigma}
	\sigma^2\triangleq \frac{2^{M+1}-2}{(1-2\eta)^2}\left(\frac{\eta (1-\eta)}{\alpha_0}+\eta^2\sigma_1^2\right).
	\end{align}
	
	In the pooled sequencing scenario, the sequencer will receive $G_n$, which is defined as 
	\begin{align} \label{mainrandom}
	G_n&=\frac{1}{\alpha_0 (1-2\eta)}\left(\sum_{m=0}^{M-1} \sum_{i=1}^{\alpha_{m,n}}\tilde{X}_{m,n,i}\right)\nonumber\\
	&+\frac{1}{\alpha_0 (1-2\eta)}\left(\sum_{k=0}^{M-1} \sum_{i=1}^{\tilde{\alpha}_{k,n}}\tilde{Y}_{k,n,i}\right)\nonumber\\
	&-\frac{\left(2^{M+1}-2\right)\eta}{1-2\eta}.
	\end{align}
	
	Consider the random variable $\sum_{i=1}^{\alpha_{m,n}} \tilde{X}_{m,n,i}$ conditioned on $X_{m,n}$. We have
	\begin{align}\label{hey}
	\sum_{i=1}^{\alpha_{m,n}} \tilde{X}_{m,n,i}&=\sum_{i=1}^{2^m\alpha_0} \tilde{X}_{m,n,i}\nonumber\\
	&+\sum_{i=2^m\alpha_0+1}^{\alpha_{m,n}} \tilde{X}_{m,n,i}.
	\end{align}
	It is trivial that the random variables $\sum_{i=1}^{2^m\alpha_0} \tilde{X}_{m,n,i}$ and $\sum_{i=2^m\alpha_0+1}^{\alpha_{m,n}} \tilde{X}_{m,n,i}$ are independent conditioned on $X_{m,n}$. Also, the distribution of $\sum_{i=2^m\alpha_0+1}^{\alpha_{m,n}} \tilde{X}_{m,n,i}$ resembles that of $\sum_{i=1}^{\alpha_{m,n}-2^m\alpha_0} \tilde{X}_{m,n,i}$ both conditioned on $X_{m,n}$.
	
	We define
	\begin{align}
	\xi_{m,n}\triangleq\alpha_{m,n}-2^m\alpha_0.
	\end{align}
	Thus we have $\mathbb{E}\left(\xi_{m,n}\right)=0$ and
	\begin{align}
	\textrm{Var}\left(\xi_{m,n}\right)=\textrm{Var}\left(\alpha_{m,n}\right)=\sigma_\alpha^2.
	\end{align}
	Similar to the steps taken in Subsection \ref{mathmode1} and as a result of the central limit theorem and orthogonality principle
	\begin{align} \label{rr3}
	&\frac{1}{\alpha_0}\sum_{i=1}^{2^m\alpha_0} \tilde{X}_{m,n,i}\ \text{conditioned on}\ X_{m,n}\nonumber\\
	&\sim\mathcal{N}\left(2^m\left((1-2\eta)X_{m,n}+\eta\right),\frac{2^m \eta(1-\eta)}{\alpha_0}\right).
	\end{align}
	For the second term in~\eqref{hey} we have
	\begin{align} \label{rr2}
	&\mathbb{E}\left(\frac{1}{\alpha_0}\sum_{i=1}^{\xi_{m,n}} \tilde{X}_{m,n,i}\ |\ X_{m,n}\right)\nonumber\\
	&=\mathbb{E}_{\xi_{m,n}}\mathbb{E}\left(\frac{1}{\bar{\alpha}}\sum_{i=1}^{\xi_{m,n}} \tilde{X}_{m,n,i}\ |\ X_{m,n},\xi_{m,n}\right)\nonumber\\
	&=\mathbb{E}_{\xi_{m,n}}\left(\frac{1}{\alpha_0}\xi_{m,n} \left((1-2\eta)X_{m,n}+\eta\right)\right)=0.
	\end{align}
	Using the law of total variance we have
	\begin{align} \label{rr1}
	&\textrm{Var}\left(\frac{1}{\alpha_0}\sum_{i=1}^{\xi_{m,n}} \tilde{X}_{m,n,i}\ |\ X_{m,n}\right)\nonumber\\
	&=\mathbb{E}_{\xi_{m,n}}\left(\textrm{Var}\left(\frac{1}{\alpha_0}\sum_{i=1}^{\xi_{m,n}} \tilde{X}_{m,n,i}\ |\ X_{m,n},\xi_{m,n}\right)\right)\nonumber\\
	&+\textrm{Var}_{\xi_{m,n}}\left(\mathbb{E}\left(\frac{1}{\alpha_0}\sum_{i=1}^{\xi_{m,n}} \tilde{X}_{m,n,i}\ |\ X_{m,n},\xi_{m,n}\right)\right)\nonumber\\
	&\stackrel{(a)}{=}\textrm{Var}_{\xi_{m,n}}\left(\frac{1}{\alpha_0}\xi_{m,n} \left((1-2\eta)X_{m,n}+\eta\right)\right)\nonumber\\
	&=\frac{((1-2\eta)X_{m,n}+\eta)^2}{(\alpha_0)^2} \textrm{Var}_{\xi_{m,n}}\left(\xi_{m,n}\right)\nonumber\\
	&=\frac{((1-2\eta)X_{m,n}+\eta)^2}{(\alpha_0)^2} \sigma_\alpha^2.
	\end{align}
	where $(a)$ results from the fact that $\mathbb{E}\left(\xi_{m,n}\right)=0$. Based on~\eqref{rr1},~\eqref{rr2},~\eqref{rr3},~\eqref{hey} and due to the MMSE rule and the orthogonality theorem we have
	\begin{align} \label{ghar1}
	\frac{1}{\alpha_0(1-2\eta)}\sum_{i=1}^{\alpha_{m,n}} \tilde{X}_{m,n,i}=\frac{\alpha_{m,n}}{\alpha_0}\left(X_{m,n}+\frac{\eta}{1-2\eta}\right)+Z_{m,n},
	\end{align}
	where $Z_{m,n}\sim\mathcal{N}\left(0,\frac{2^m\eta(1-\eta)}{\alpha_0(1-2\eta)^2}\right)$. Using the same steps, for the data collector we have
	\begin{align} \label{ghar2}
	\frac{1}{\alpha_0(1-2\eta)}\sum_{i=1}^{\tilde{\alpha}_{k,n}} \tilde{Y}_{k,n,i}=\frac{\tilde{\alpha}_{k,n}}{\alpha_0}\left(y_{k,n}+\frac{\eta}{1-2\eta}\right)+\tilde{Z}_{k,n},
	\end{align}
	where $\tilde{Z}_{k,n}\sim\mathcal{N}\left(0,\frac{2^k\eta(1-\eta)}{\alpha_0(1-2\eta)^2}\right)$.
	
	Therefore using~\eqref{ghar1} and~\eqref{ghar2},~\eqref{mainrandom} can be written as
	\begin{align}
	G_n&=\sum_{m=0}^{M-1}\frac{\alpha_{m,n}}{\alpha_0}\left(X_{m,n}+\frac{\eta}{1-2\eta}\right)\nonumber\\
	&+\sum_{k=0}^{M-1}\frac{\tilde{\alpha}_{k,n}}{\alpha_0}\left(y_{k,n}+\frac{\eta}{1-2\eta}\right)\nonumber\\
	&-\frac{\left(2^{M+1}-2\right)\eta}{1-2\eta}+Z'_n, \label{mainrand}
	\end{align} 
	where
	\begin{align}
	Z'_{n}\triangleq\sum_{m=0}^{M-1}\left(Z_{m,n}+\tilde{Z}_{m,n}\right).
	\end{align}
	Thus
	\begin{align}
	Z'_{n}\sim\mathcal{N}\left(0,\frac{2^{M+1}-2}{\alpha_0}\frac{\eta (1-\eta)}{(1-2\eta)^2}\right)
	\end{align}
	For the fraction $\frac{\alpha_{m,n}}{\alpha_0}$ we can write it as
	\begin{align} \label{72}
	\frac{\alpha_{m,n}}{\alpha_0}=2^m+\frac{\zeta_{m,n}}{\alpha_0},
	\end{align}
	where
	\begin{align}
	\zeta_{m,n}\triangleq \alpha_{m,n}-2^m\alpha_0.
	\end{align}
	Therefore $\textrm{Var}(\zeta_{m,n})=\textrm{Var}\left(\alpha_{m,n}\right)$ and for $\delta_{m,n}\triangleq \frac{\zeta_{m,n}}{\alpha_0}$ we have
	\begin{align} \label{74}
	\textrm{Var}\left(\delta_{m,n}\right)=\frac{\textrm{Var}\left(\alpha_{m,n}\right)}{(\alpha_0)^2}=2^m\sigma_1^2.
	\end{align}
	Also, $\tilde{\delta}_{k,n}$ is defined similarly. Using~\eqref{mainrand},~\eqref{72}, and~\eqref{74},~\eqref{sss} is resulted from~\eqref{mainrandom}
	
	\subsection{Estimation Rule} \label{est}
	For any SNP position $n\in [N]$, the objective for the data collector is to estimate the vector $\mathbf{x}_n=\left[X_{1,n}, X_{2,n},\cdots, X_{M,n}\right]^T$. We define the extended vector $\tilde{\mathbf{x}}_n\triangleq\left[X_{1,n},\cdots,X_{M,n},y_{1,n},\cdots,y_{K,n}\right]^T$, where the last $K$ entries are known to the data collector. Therefore, for the data collector, estimating $\tilde{\mathbf{x}}_n$ is equivalent to estimating $\mathbf{x}_n$.
	
	In this section, our objective is to find the rule that should be used by the data collector to estimate $\tilde{\mathbf{x}}_n$. Using the ML rule, the estimate $\hat{\tilde{\mathbf{x}}}_n$ is obtained by
	\begin{align}
	\hat{\tilde{\mathbf{x}}}_n&=\arg\max_{\tilde{\mathbf{x}}_n} \mathbb{P}(G_n\ |\ \tilde{\mathbf{x}}_n)\nonumber\\
	&=\arg\max_{\tilde{\mathbf{x}}_n} \mathbb{P}\left(G_n-\sum_{m=0}^{M-1}2^m(X_{m,n}+y_{m,n})\ |\ \tilde{\mathbf{x}}_n\right),
	\label{MLeqn}
	\end{align}
	Let
	\begin{align}
	V_n\triangleq G_n-\sum_{m=0}^{M-1}2^m(X_{m,n}+y_{m,n}).
	\end{align}
	Based on~\eqref{sss},
	\begin{align}
	V_n\ \text{conditioned on}\ \tilde{\mathbf{x}}_n\sim \mathcal{N}\left(0,(2^{M+1}-2)\sigma_1^2+\sigma^2\right).
	\end{align}
	Therefore,
	\begin{align} \label{estRule}
	\hat{\tilde{\mathbf{x}}}_n=\arg\min_{\tilde{\mathbf{x}}_n} |V_n|.
	\end{align}
	
	\subsection{Proof of Theorem \ref{thm3}} \label{thm3proof}
	Based on the mathematical model and estimation rule presented in the previous subsections, we are ready to provide the proof of theorem \ref{thm3}. 
	\begin{proof}
		Similar to the proof presented in subsection \ref{proof1} and based on the estimation rule in~\eqref{estRule}, our estimation resembles an AWGN channel. In other words, if we estimate $\sum_{m=0}^{M-1}2^m(X_{m,n}+y_{m,n})$, then $\hat{\tilde{\mathbf{x}}}_n$ is resulted accordingly.
		Thus, for the probability of error we have
		\begin{align}
		\mathbb{P}(\text{error}) &\leq Q\left(\frac{1}{2\sqrt{(2^{M+1}-2)\sigma_1^2+\sigma^2}}\right)\nonumber\\
		&\leq\exp\left(\frac{-1}{8\left((2^{M+1}-2)\sigma_1^2+\sigma^2\right)}\right).
		\end{align}
		Putting the right-hand side less than $\epsilon$ results
		\begin{align} \label{70}
		(2^{M+1}-2)\sigma_1^2+\sigma^2\leq\frac{1}{8\ln\left(\frac{1}{\epsilon}\right)}.
		\end{align}
		Rewriting the left-hand side by substituting $\sigma_1^2$ results
		\begin{align}
		&(2^{M+1}-2)\sigma_1^2+\sigma^2=\nonumber\\
		&\frac{2^{M+1}-2}{(1-2\eta)^2}\left(\frac{\eta (1-\eta)}{\alpha_0}+(\eta^2+(1-2\eta)^2)\sigma_1^2\right).
		\end{align}
		In order~\eqref{70} to hold, it is sufficient for both two terms in the right-hand side of the above equality to be less than $\frac{1}{16\ln\left(\frac{1}{\epsilon}\right)}$. From the first inequality we have
		\begin{align}
		\alpha_0\geq \frac{16\eta (1-\eta)}{(1-2\eta)^2}\left(2^{M+1}-2\right)\ln\left(\frac{1}{\epsilon}\right).
		\end{align}
		From the second one we reach
		\begin{align}
		\alpha_0\geq \sqrt{16\sigma_\alpha^2\left(1+\frac{\eta^2}{(1-2\eta)^2}\right)\left(2^{M+1}-2\right)\ln\left(\frac{1}{\epsilon}\right)}.
		\end{align}
		As both inequalities above should hold, the theorem is proven.
		
	\end{proof}
	
	\subsection{Discussion} \label{disc}
	First of all, if we put $\sigma_\alpha=0$ in Theorem \ref{thm3}, the result resembles that of Theorem \ref{thm1} which was expected. Also, as it is seen from \ref{thm3}, by increasing $M$ and decreasing $\epsilon$, $e_1$ grows much faster (quadratic) than $e_2$. So for small enough $\sigma_\alpha$, $e_1$ is probably the bigger value.
	\begin{remark}
		In this remark we will show that theorem \ref{thm2} works in the random case of coverage depth too. Similar to the steps taken in the Subsection \ref{mathmodel}, the sequencer will receive $q_n$ in SNP position $n\in [N]$ such that
		\begin{align} \label{ssss}
		q_n=&\sum_{m=0}^{M-1} \left(2^m+\delta_{m,n}\right)X_{m,n}\nonumber\\
		+&\sum_{k=0}^{M-1} \left(2^k+\tilde{\delta}_{k,n}\right)Y_{k,n} + \tilde{Z}_n
		\end{align}
		where $\delta_{m,n}$ and $\tilde{\delta}_{k,n}$ are normal random variables with zero mean and variance $\sigma_1^2$. Also, $\tilde{Z}_n\sim\mathcal{N}(0,\sigma^2)$ where $\sigma^2$ is defined in~\eqref{sigma}.
		We can write $q_n$ as
		\begin{align} \label{sssss}
		q_n=\sum_{m=0}^{M-1}2^m(X_{m,n}+Y_{m,n})+\hat{Z}_n,
		\end{align}
		where $\hat{Z}_n\sim\mathcal{N}\left(0,(2^{M+1}-2)\sigma_1^2+\sigma^2\right)$.
		
		From~\eqref{sssss} the Markov chain in the proof of Theorem \ref{thm2} is valid here as well. Continuing the same steps, we conclude that Theorem \ref{thm2} works here. Therefore, all the discussions in that scenario is still valid here.
	\end{remark}
	\begin{remark}
		All the results driven are for the case of Haploid cells. In this case, there is one set of chromosomes. But in the case of Diploid cells, each cell carries two sets of chromosomes. It means that in every position in the genome, there are two chromosomes covering it. To extend our results to the case of Diploid cells, we can assume each individual contains the chromosomes of two haploid-celled individuals. So all the results are tailored to the case of Diploid cells if we replace $M$ with $2M$ for the $M$-individual scheme.
	\end{remark}
	
	\section{Conclusion and Future Steps} \label{sec6}
	In this paper, we introduced the problem of privacy in the process of DNA sequencing. Previously, the privacy criterion was inspected in genomic data sets, but their concern of privacy is very different in comparison to our perspective. We seek to satisfy privacy in the process of sequencing that enables to hide the DNA sequence from the sequencing machine, while letting us to construct the sequence in a local processor that is trusted. Previous approaches' concern was briefly how to make genomic data ready for announcement in a way that the information of no single individual is violated, so one can see how our approach is different.
	
	In this paper, we aimed to theoretically define the problem of \textit{privacy in DNA sequencing} and introduce an achievable scheme so that it can satisfy our constraints if parameters are adjusted correctly. We used non-colluding sequencers and distributed the genome data between them. Also, we used the idea of pooled sequencing and combined our the real data with \textit{known sequences}. By setting the number of known sequences and the coverage depth of sequences, we can satisfy the constraints.
	
	As this is the first paper in this problem, there can be done a lot in future works. For instance, The case in which a set of sequencers are collaborating could be concerned, or the case in which fragments are not limited to contain just one SNP. Also, the lower bounds in the theorems in this paper can be improved. At last, we hope this paper has paved the way towards privacy in the process of sequencing.
	
	\bibliographystyle{ieeetr}
	\bibliography{journal_abbr,CodDecB}
	
	\newpage
	
	\vfill

\end{document}